\documentclass[10pt]{iopart}
\expandafter\let\csname equation*\endcsname\relax
\expandafter\let\csname endequation*\endcsname\relax
\usepackage{amsmath}
\usepackage{psfrag}
\usepackage{graphicx}
\usepackage{graphics}
\usepackage{epsfig}
\usepackage{bm}
\usepackage{color}
\usepackage{verbatim,color}
\usepackage{physics}
\usepackage[acronym]{glossaries}
\usepackage[normalem]{ulem}
\usepackage{mathrsfs}
\usepackage{mathtools}
\usepackage{multirow}
\usepackage[overload]{empheq}
\usepackage{subfig}

\newacronym{dr}{DR}{dimensional regularization}
\newacronym{ms}{MS}{minimal subtraction}
\newacronym{eft}{EFT}{effective field theory}
\newacronym{mqst}{MQST}{macroscopic quantum self-trapping }
\newacronym{lhy}{LHY}{Lee-Huang-Yang}
\newcommand{\beq}{\begin{equation}}
\newcommand{\eeq}{\end{equation}}
\newcommand{\beqa}{\begin{eqnarray}}
\newcommand{\eeqa}{\end{eqnarray}}
\newcommand{\ba}{\begin{aligned}[b]}
\newcommand{\ea}{\end{aligned}}
\newcommand{\cblue}{\color{black}}

\begin{document}

\title{Quantum fluctuations in atomic Josephson junctions: the role of dimensionality}

\author{A. Bardin$^{1, 2}$, F. Lorenzi$^{1,2}$, and L. Salasnich$^{1,2,3,4}$}

\address{$^{1}$Dipartimento di Fisica e Astronomia ``Galileo Galilei", 
Universit\`a di Padova, Via Marzolo 8, 35131 Padova, Italy\\
$^{2}$Istituto Nazionale di Fisica Nucleare (INFN), Sezione di Padova, via Marzolo 8, 35131 Padova, Italy\\
$^{3}$Padua QTech Center, Universit\`a di Padova, 
via Gradenigo 6A, 35131 Padova, Italy\\
$^{4}$Istituto Nazionale di Ottica (INO) del Consiglio Nazionale delle Ricerche (CNR), via Nello Carrara 1, 50019 Sesto Fiorentino, Italy
}

\begin{abstract}
    We investigate the role of quantum fluctuations in the dynamics of a bosonic Josephson junction in $D$ spatial dimensions, by using beyond-mean-field Gaussian corrections. We derive some key dynamical properties in a systematic way for $D=3, 2, 1$. 
    In particular, we compute the Josephson frequency in the regime of low population imbalance. We also obtain the critical strength of the macroscopic quantum self-trapping. Our results show that quantum corrections increase the Josephson frequency in spatial dimensions $D=2$ and $D=3$, but they decrease it in the $D=1$ case. The critical strength of macroscopic quantum self-trapping is instead reduced by quantum fluctuations in $D=2$ and $D=3$ cases, while it is enhanced in the $D=1$ configuration. {\cblue We show that the difference between the cases of $D=2$ and $D=3$ on one side, and $D=1$ on the other, can be related to the qualitatively different dependence of the interaction strength on the scattering length in the different dimensions.}
\end{abstract}

\maketitle

\section{Introduction}
A Josephson junction is a device composed of a superconductor, or a superfluid, with a tunneling barrier separating two regions. While the first studies on this kind of problem were targeting superconductors \cite{josephson1962possible}, the achievement of Bose-Einstein condensation prompted investigations on this model in the setting of ultracold atomic systems too \cite{smerzi1997quantum}. {\cblue Experiments using bosonic Josephson junctions have been performed sice 2005 \cite{albiez2005direct} and also ac and dc Josephson effects have been measured \cite{levy2007acdc}}. In contrast to superconducting Josephson junctions, atomic Josephson junctions can exhibit a significant difference in population between the two sides, and a self-trapping phenomenon called \gls{mqst} {\cblue \cite{raghavan1999coherent}. Eventually, the presence of normal currents has been shown to eliminate the MQST regime in long evolution times \cite{meier2001tunneling}}. 
In the context of atomic Josephson junctions, the role of dimensionality has been studied in the context of elongated 1-dimensional junctions \cite{ji2022floquet} consisting in two sites spatially separated by an optical potential. {\cblue Experimental realizations of the atomic Josephson junction can be based on the implementation of a narrow potential barrier that spatially separate the condensate, and allows tunnelling. In this scenario, it is possible study extended Josephson junctions in $D=1$ \cite{pigneur}, corresponding to a potential that confines the condensate around two parallel axis, and $D=2$, in the case of separation in two parallel planes. This approach is called external coupling \cite{farolfi2021manipulation}. In opposition, the technique of internal coupling has beed recently shown to be promising: two internal states in the condensate are coupled by Rabi resonant radiation, thus obtaining an effective Josephson junction without the implementation of a narrow potential \cite{farolfi2021manipulation, fava2018observation}. In the latter case, it is possible to investigate the behaviour of the junction also in $D=3$, a case that was sorted out with external coupling due to the need of the barrier. Realizing extended atomic Josephson junctions is more difficult with respect to single-mode systems, because of the presence of inhomogeneities in the atom density and external fields. Analyzing the full dyamics at the Gross-Pitaevskii level, one obtains rich phase diagram including dissipative, self-trapped regimes \cite{xhani2020dynamical}. Analysis with $D=1$ \cite{polo2021diagram, saha2021} and $D=2$ \cite{momme2019collective, singh2020two} have highlighted the emergence of novel dissipation mechanisms, dispersive shock waves and BKT transition.}
Moreover, the tunability of the interaction strength via Feshbach resonances allowed investigations with fermionic superfluids near the BEC-BCS crossover, in experimental \cite{valtolina2015josephson,delpace2021transpost,kwon2020strongly,burchianti2018connecting} and theoretical \cite{spuntarelli2007josephson,zaccanti2019critical,wlazlowski2023dissipation,piselli2020finite} studies. In the case of fermionic superfluid, the role of dimensionality has been experimentally investigated in the $D=2$ case \cite{luick2020ideal}.
Another interesting setup, made available using optical lattices, is the realization of many Josephson junctions in an array shape \cite{cataliotti2001josephson}.
Such kind of system had been proposed as a platform for quantum computing \cite{tian2003quantum}.

Typically, the complete quantum dynamics of Josephson junctions is explained using the phase model, which depends on the commutation rule of the phase operator $\hat{\phi}$ with the number operator $\hat{N}$ \cite{leggett1991concept}. {\cblue By looking at the population imbalance and at the relative phase between the two sites, is it possible to separate two qualitatively different dynamical regimes: the first one is the Josephson regime, which is characterized by sinusoidal oscillations of both quantity around a zero mean value (in the deep tunneling regime). The oscillation frequency is called the Josephson frequency. In the second regime, {\cblue the self-trapping one}, the population imbalance exhibits small amplitudes oscillation around a non-zero mean value while the relative phase increases in time. The rate of increase of the phase with time is called phase-slippage rate \cite{avenel}. The phase model} approach is the starting point of many theoretical studies of Josephson junctions. For example, the effect of the finite size of the system had been tackled by using the so-called atomic coherent states \cite{wimberger2021finite}. In this case, the computation of corrections to the \gls{mqst} critical strength was shown to be particularly subtle. In another study, the path integration technique had been used for obtaining an effective action depending only on the phase dynamical variable: this approach was used to compute quantum corrections to the Josephson frequency that are in principle verifiable for both atomic and superconducting systems in specific regimes \cite{furutani2022quantum}.

In the present work, we first review, in Section~\ref{sec:MF} the mean-field calculations of the Josephson frequency and the \gls{mqst} critical strength, serving as a starting point for our analysis. {\cblue In Sections~\ref{sec:3d}, \ref{sec:1d} and~\ref{sec:2d},} by including beyond-mean-field Gaussian corrections in $D$ spatial dimensions \cite{sala-zero} on each of the two sites, we systematically calculate the effect of quantum fluctuations for the case, respectively, of {\cblue $D=3, 1, 2$}. These quantum fluctuations give rise to the Lee-Huang-Yang correction \cite{lhy} for $D=3$, to the Schick-Popov correction \cite{schick,popov} for $D=2$, and to the next-to-leading term of the Lieb-Liniger theory \cite{lieb} for $D=1$. {\cblue In Section \ref{sec:slip} we compare the phase-slippage rates in the self-trapping regimes, by solving numerically the equations of motion.}
Our approach provides an analytically treatable approximation of the full quantum dynamics, and significant corrections are derived.

\section{Mean-field results}\label{sec:MF}

\subsection{D-dimensional case}
We want to construct an effective Lagrangian for a bosonic system with two sites of volume $V=L^D$ each \cite{furutani2022quantum}. The corresponding Lagrangian density is made of three terms:
\begin{equation}
    \mathscr{L}=\mathscr{L}_1+\mathscr{L}_2+\mathscr{L}_{J}.
\end{equation}
The first and the second term are given by
\begin{equation}
    \mathscr{L}_k=i\hbar\Phi_k^*(t)\partial_t\Phi_k(t)-\frac{1}{2}g_0|\Phi_k(t)|^4\quad\quad k=1,2,
\end{equation}
where $\Phi_k(t)$ is a complex time-dependent field describing the bosons in one of the sites ($k=1,2$) and $g_0$ is the coupling constant. It is important to stress that the spatial dependence of the 
field $\Phi_k(t)$ is encoded only in the subindex $k$. {\cblue If we assume  that the wavefunctions of the two sites have small overlapping in the barrier region, we can write} the third term, which phenomenologically introduces tunneling (hopping), {\cblue as}
\begin{equation}
    \mathscr{L}_{J}=\frac{J}{2}\left(\Phi_1^*(t)\Phi_2(t)+\Phi_2^*(t)\Phi_1(t)\right) , 
\end{equation}
the constant $J$ is connected to the exchange of particles between the two sites. Integrating in space the Lagrangian density one obtains the Lagrangian \cite{furutani2022quantum}
\begin{equation}
\begin{split}
    \mathcal{L}&=\int_V \mathscr{L}\ \dd[D]{\Vec{r}} = L^D\mathscr{L} \\    &=\sum_k\Bigl(i\hbar\varphi^*_k(t)\partial_t\varphi_k(t)-\frac{U}{2}|\varphi_k(t)|^4\Bigr) +\frac{J}{2}(\varphi_1^*(t)\varphi_2(t)+\varphi_2^*(t)\varphi_1(t)),
\end{split}
\end{equation}
where the new renormalized functions describing the system are
\begin{equation}
    \varphi_k(t)\equiv\sqrt{L^D}\Phi_k(t)\quad k=1,2 \ ,
\end{equation}
and 
\begin{equation}
    U\equiv\frac{g_0}{L^D}.
\end{equation}
Through the Madelung transformation \cite{madelung1927quantentheorie}, given by
\begin{equation}
    \varphi_k(t)=\sqrt{N_k(t)}e^{i\phi_k(t)}\quad\quad k=1,2 \ ,
    \label{madelung}
\end{equation}
the complex function describing the bosons can be rewritten in terms of its phase $\phi_k(t)$ and its modulus $\sqrt{N_k(t)}$, where the square of the latter corresponds to the number of bosons in the $k-$th site.
Then the Lagrangian becomes
\begin{equation}
\begin{split}
    \mathcal{L}=&\sum_k\Bigl(i\hbar\frac{\Dot{N}_k}{2}-\hbar\Dot{\phi}_kN_k-\frac{U}{2}N_k^2\Bigr)+J\cos{(\phi_1-\phi_2)}\sqrt{N_1N_2}.
\end{split}
\end{equation}
Introducing the total number of particles $N\equiv N_1(t)+N_2(t)$, the relative phase $\phi(t)\equiv\phi_2(t)-\phi_1(t)$, the total phase $\Bar{\phi}(t)\equiv\phi_1(t)+\phi_2(t)$ and the population imbalance $z(t)\equiv (N_1(t)-N_2(t))/N$ then
\begin{equation}
\begin{split}
        \mathcal{L}=&i\hbar\frac{\Dot{N}}{2}+\frac{N\hbar}{2}z\Dot{\phi}-\frac{N\hbar}{2}\Dot{\Bar{\phi}}-\frac{UN^2}{4}-\frac{UN^2}{4}z^2 +\frac{JN}{2}\sqrt{1-z^2}\cos{\phi},
\end{split}
\end{equation}
however, the first term is zero since $N$ is constant and the third and fourth terms can be removed since the former is  an exact differential and the latter is constant. The Lagrangian then reduces to
\begin{equation}
    \mathcal{L}=\frac{N\hbar}{2}z\Dot{\phi}-\frac{UN^2}{4}z^2+\frac{JN}{2}\sqrt{1-z^2}\cos{\phi},
    \label{Lagrangian}
\end{equation}
and the corresponding Euler-Lagrange equations, called Josephson-Smerzi equations \cite{smerzi1997quantum} are
\begin{equation}
    \begin{cases}
    \Dot{z}=-\frac{J}{\hbar}\sqrt{1-z^2}\sin{\phi}\\
    \Dot{\phi}=\frac{J}{\hbar}\frac{z}{\sqrt{1-z^2}}\cos{\phi}+\frac{UNz}{\hbar}.
    \end{cases}
    \label{ELEQ}
\end{equation}

\subsection{Josephson frequency}
To obtain a quadratic Lagrangian one consider the limit in which $|\phi(t)|\ll1$ and $|z(t)|\ll1$. 
Therefore the linearized Josephson-Smerzi equations are 
\begin{equation}
     \begin{cases}
    \Dot{z}=-\frac{J}{\hbar}\phi\\
    \Dot{\phi}=\frac{J+UN}{\hbar}z,
    \end{cases}
\end{equation}  
and from these equations, one can get the harmonic oscillator equations for the population imbalance $z(t)$ and the relative phase $\phi(t)$:
\begin{equation}
    \begin{cases}
        \Ddot{z}+\Omega_{mf}^2z=0\\
        \Ddot{\phi}+\Omega_{mf}^2\phi=0,
    \end{cases}
\end{equation}
where the Josephson frequency is introduced \cite{smerzi1997quantum}, its expression is given by
    \begin{equation}
        \Omega_{mf}=\frac{1}{\hbar}\sqrt{J^2+UNJ} = \frac{1}{\hbar}\sqrt{J^2+Jg_0n}.
        \label{Omega}
    \end{equation}
Note that there are two particular regimes. If $J\gg UN$ then the frequency can be approximated with the Rabi frequency $\Omega_R$ \cite{smerzi1997quantum}
\begin{equation}
    \Omega_{mf}\simeq\Omega_R=\frac{J}{\hbar}.
\end{equation}
Vice versa, if $J\ll UN$ then the frequency can be approximated to \cite{smerzi1997quantum}
\begin{equation}
    \Omega_{mf}\simeq\Omega_J=\frac{\sqrt{UNJ}}{\hbar} .
\end{equation}
Hence the frequency ($\ref{Omega}$) can be rewritten as a function of these two particular cases as
\begin{equation}
    \Omega_{mf}=\sqrt{\Omega^2_R+\Omega^2_J}.
\end{equation}

\subsection{Macroscopic Quantum Self Trapping}
In order to describe the \gls{mqst} in the mean-field approximation, one needs to find the conserved energy of the system, which is given by
\begin{equation}
    E=\Dot{\phi}\frac{\partial \mathcal{L}}{\partial \Dot{\phi}}-\mathcal{L},
\end{equation}
however the Lagrangian is independent from $\Dot{z}$, %
therefore the conserved energy is given by
\begin{equation}
    E=\frac{UN^2}{4}z^2-\frac{JN}{2}\sqrt{1-z^2}\cos{\phi}.
    \label{E}
\end{equation}
The \gls{mqst} happens when $\langle z\rangle\neq0$ and the condition to have \gls{mqst} is given, calling $z_0=z(0)$ and $\phi_0=\phi(0)$, by the following inequality
\begin{equation}
    E(z_0,\phi_0)>E(0,\pi),
    \label{INEQ}
\end{equation}
since $z(t)$ cannot become zero during an oscillation cycle.
The MQST condition can be expressed also with a dimensionless parameter, known as strength, defined as
\begin{equation}
    \Lambda\equiv\frac{NU}{J}.
    \label{Xi}
\end{equation}
In fact, inserting (\ref{E}) and (\ref{Xi}) into (\ref{INEQ}) , one has 
\begin{equation}\label{Xi-criticalMF}
    \Lambda>\Lambda_{\text{c,\ mf}} {\cblue \equiv\frac{1+\sqrt{1-z_0^2}\cos{\phi_0}}{z_0^2/2}.}
\end{equation}
where we defined the critical value of the strength, above which the \gls{mqst} occurs.
Eq. (\ref{Xi-criticalMF}) is the familiar mean-field condition to achieve \gls{mqst} in Bose-Einstein condensates.

In the next sections, the results found in the mean-field approximation are used as a starting point to compute the beyond-mean-field corrections to the Josephson frequency and the \gls{mqst}. 
As we will see, contrary to the mean-field results, the beyond-mean-field Gaussian corrections are strongly dependent on the spatial dimension $D$. 

\begin{section}{Beyond-mean-field: D=3 case}\label{sec:3d}

We can express the beyond-mean-field energy density in $D=3$ as \cite{sala-zero,lhy}
\begin{equation}
    \mathcal{E}=\frac{1}{2}g_0n^2+\frac{8}{15\pi^2}\sqrt{\frac{m}{\hbar^2}}^3(g_0n)^{\frac{5}{2}},
    \label{E3D}
\end{equation}
{\cblue where in the $D=3$ case the coupling constant $g_0$ depends on the s-wave scattering length $a_s$ by the following relation
\begin{equation}
    g_0=\frac{4\pi\hbar^2a_s}{m}.
\end{equation}
The beyond-mean-field energy density is obtained from the beyond-mean-field grand potential. In particular, in a generic spatial dimension $D$, the latter is calculated renormalizing the integral over the momentum-space of the Bogoliubov spectrum, namely:
\begin{equation}
     \frac{1}{2}\int \dd[D]{\mathbf{q}} \ E_{\mathbf{q}}(\mu)=\frac{1}{2}\int \dd[D]{\mathbf{q}} \ \sqrt{\frac{\hbar^2q^2}{2m}\left(\frac{\hbar^2q^2}{2m}+2\mu\right)},
     \label{Bogoliubov}
\end{equation}
where $\mu$ is the chemical potential and $\mathbf{q}$ is the linear momentum. This integral is the zero-point energy of the excitations, which exhibit an ultraviolet divergency, but we can regularize it by means of dimensional regularization \cite{sala-zero}. In this section we consider $D=3$. Once one has the beyond-mean-field correction term to the grand potential, it is possible to retrieve the beyond-mean-field energy density as follow: firstly, one calculates the beyond-mean-field number density performing a derivative with respect to the chemical potential to the beyond-mean-field grand potential. Then, inverting the obtained number density relation, one obtains an expression for the beyond-mean-field chemical potential. Finally, integrating the latter expression in the number density one retrieves the beyond-mean-field energy density. Considering also (\ref{E3D}),
} the Lagrangian density acquires a new term
\begin{equation}
    \begin{split}
    \mathscr{L}=&\mathscr{L}_0-\frac{8g_0^{\frac{5}{2}}\sqrt{m^3}}{15\pi^2\hbar^3}\left(|\Phi_1(t)|^5+|\Phi_2(t)|^5\right),
    \end{split}
\end{equation}
where $\mathscr{L}_0$ is the mean-field Lagrangian density.
Integrating the Lagrangian density in space one obtains
\begin{equation}
\begin{split}
    \mathcal{L}&=\mathcal{L}_0-\frac{8g_0^{\frac{5}{2}}\sqrt{m^3}}{15\pi^2\hbar^3L^\frac{9}{2}}\left(|\varphi_1(t)|^5+|\varphi_2(t)|^5\right).
\end{split}
\end{equation}

\begin{subsection}{Josephson frequency}
Performing the Madelung transformation (\ref{madelung}) one obtains
\begin{equation}
        \mathcal{L}=\mathcal{L}_0-\frac{8\sqrt{m^3g_0^5}}{15\pi^2\hbar^3L^{\frac{9}{2}}}\left(N_1^\frac{5}{2}(t)+N_2^\frac{5}{2}(t)\right),
\end{equation}
and rewriting the number of particles in each site as a function of the total number of particles and the population imbalance $N_{1,2} = N(1\pm z)/2$ one gets
\begin{equation}
    \mathcal{L}=\mathcal{L}_0-\frac{\sqrt{2m^3g_0^5}N^{\frac{5}{2}}}{15\pi^2\hbar^3L^{\frac{9}{2}}}\left[(1+z)^{\frac{5}{2}}+(1-z)^{\frac{5}{2}}\right].
    \label{3DLBMF0}
\end{equation}
Now, since we are in the low population imbalance limit, it is possible to do the following expansion
\begin{equation}
    (1\pm z)^{n}\simeq1\pm nz+\frac{n(n-1)z^2}{2},
    \label{expansion1}
    \end{equation}
and summing the two contributions
\begin{equation}
    (1+z)^{n}+(1-z)^{n}\simeq2+n(n-1)z^2.
\end{equation}
Inserting it into (\ref{3DLBMF0}) and removing the term constant in $z$ stemming from the calculations, one obtains
    \begin{equation}
        \mathcal{L}=\mathcal{L}_0-\frac{\sqrt{m^3g_0^5}N^\frac{5}{2}}{2\sqrt{2}\pi^2\hbar^3L^{\frac{9}{2}}}z^2.
        \end{equation}
    Finally, the Lagrangian in the $D=3$ case is
        \begin{equation}
        \begin{split}
          \mathcal{L}=&\frac{N\hbar }{2}z\Dot{\phi}-\left(\frac{UN^2+JN}{4}\right)z^2 -\frac{JN}{4}\phi^2-\frac{\sqrt{m^3g_0^5}N^\frac{5}{2}}{2\sqrt{2}\pi^2\hbar^3L^{\frac{9}{2}}}z^2 .
        \end{split}
        \end{equation}
        \\
    The Euler-Lagrange equations are
    \begin{equation}\label{EL-freq}
        \begin{cases}
            \ddot{\phi}+\Omega^2\phi=0\\
            \ddot{z}+\Omega^2z=0,
        \end{cases}
    \end{equation}
where the corrected Josephson frequency is 
\begin{equation}
    \Omega\equiv\frac{1}{\hbar}\sqrt{J^2+JUN+\frac{J\sqrt{2g_0^5n^3m^3}}{\pi^2\hbar^3}},
\end{equation}
Defining the reference energy $\varepsilon_s$ and the gas parameter $\gamma$  as
\begin{equation}
    \varepsilon_s\equiv\frac{\hbar^2}{ma_s^2}\qquad\gamma\equiv a^3_sn,
\end{equation}
the Josephson frequency can also be written, using the definition of Rabi frequency $\Omega_R$, as 
\begin{equation}
    \Omega=\Omega_R\sqrt{1+4\pi\gamma\frac{\varepsilon_s}{J}\left(1+8\sqrt{\frac{2\gamma}{\pi}}\right)}.
\end{equation}
To understand the magnitude of the beyond-mean-field correction to the Josephson frequency the ratio between the beyond-mean-field Josephson frequency $\Omega$ and the mean-field one $\Omega_{mf}$ as a function of the strength parameter, given by $\Lambda=4\pi\gamma\varepsilon_s/J$, is done. Namely it is considered
\begin{equation}
    \frac{\Omega}{\Omega_{mf}}=\sqrt{\frac{1+\Lambda\left(1+8\sqrt{\frac{2\gamma}{\pi}}\right)}{1+\Lambda}}.
\end{equation}
Looking at Fig.~\ref{fig:3DO} one observes the following behavior: the correction is more significant at higher strength parameters $\Lambda$. For strength parameters $\Lambda\rightarrow0$ the beyond-mean-field correction is irrelevant regardless of the gas parameter. Instead, for larger $\Lambda$, the relative correction is given by
\begin{equation}
    \frac{\Omega}{\Omega_ {mf}}\Bigg|_{\Lambda\gg1}=\sqrt{1+8\sqrt{\frac{2\gamma}{\pi}}}.
\end{equation}
Focusing now on the bounds of the gas parameter $\gamma$, while the lower bound is $\gamma=0$ and this is given by the fact that both the quantities defining $\gamma$, namely the s-wave scattering length $a_s$ and the number density $n$ are non-negative quantities. Instead, the upper bound limit is due to the fact that to obtain the beyond-mean-field correction we used a perturbative approach, assuming $\gamma\ll1$, for this reason, by setting a beyond-mean-field relative correction to the number density smaller than $0.1$, it follows that the upper bound on $\gamma$ must be set to $\gamma=3\times10^{-4}$, as reported in Fig.~\ref{fig:3DO}. In Fig.~\ref{fig:3DO} the relative correction is higher for larger values of the gas parameter, while for $\gamma=0$ one retrieves the mean-field case.
\begin{figure}
\centering
\includegraphics[width=0.6\textwidth]{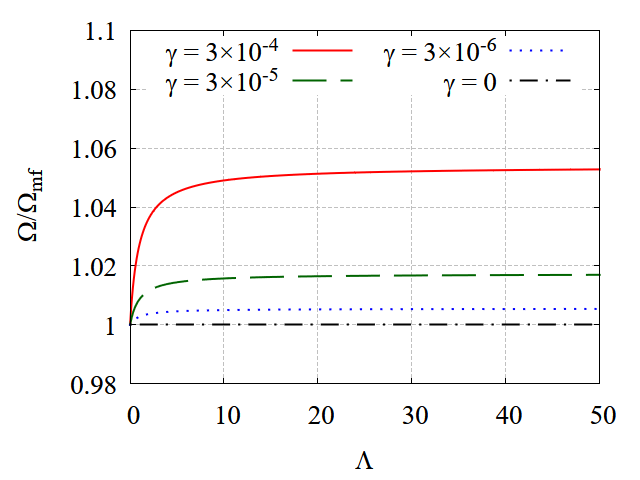}
\caption{3D beyond-mean-field relative correction to the Josephson frequency.\\ In the plot is pictured the ratio between the beyond-mean-field Josephson frequency $\Omega$ and the mean-field one $\Omega_ {mf}$ as a function of the strength parameter $\Lambda=g_0n/J$ for different values of the gas parameter $\gamma=a^3_sn$: $\gamma=3\times10^{-4}$ (red solid line),
$\gamma=3\times10^{-5}$ (green dashed line)
$\gamma=3\times10^{-6}$ (blue dotted line) and $\gamma=0$ (black dash-dotted line). The last line corresponds to the mean-field case.
}
\label{fig:3DO}
\end{figure}
\end{subsection}

\begin{subsection}{Macroscopic Quantum Self Trapping}
Starting from the beyond-mean-field Lagrangian written in terms of the total number of particles $N$, the population imbalance $z$, and the phase difference $\phi$
\begin{equation}
\begin{split}
        \mathcal{L}=&\frac{N\hbar}{2}z\Dot{\phi}-\frac{UN^2}{4}z^2+\frac{JN}{2}\sqrt{1-z^2}\cos{\phi}\\&-\frac{\sqrt{2m^3g_0^5}N^{\frac{5}{2}}}{15\pi^2\hbar^3L^{\frac{9}{2}}}\left[(1+z)^{\frac{5}{2}}+(1-z)^{\frac{5}{2}}\right],
\end{split}
\end{equation}
one finds that the conserved energy is
\begin{equation}
\begin{split}
    E=&\frac{UN^2}{4}z^2-\frac{JN}{2}\sqrt{1-z^2}\cos{\phi} +\frac{L^3\sqrt{2m^3}}{15\pi^2\hbar^3}(UN)^{\frac{5}{2}}\left[(1+z)^{\frac{5}{2}}+(1-z)^{\frac{5}{2}}\right].
\end{split}
\end{equation}
Imposing the inequality condition (\ref{INEQ}) to have \gls{mqst} one gets
\begin{equation}
\begin{split}
    &\frac{\Lambda}{2}z^2_0-\sqrt{1-z^2_0}\cos{\phi_0}+\frac{2L^3\sqrt{2m^3}}{15\pi^2\hbar^3}\Lambda U^{\frac{3}{2}}N^{\frac{1}{2}}\left[(1+z_0)^{\frac{5}{2}}+(1-z_0)^{\frac{5}{2}}\right] \\&>1+\frac{4L^3\sqrt{2m^3}}{15\pi^2\hbar^3}\Lambda U^\frac{3}{2}N^\frac{1}{2},
\end{split}
\end{equation}
and finally
\begin{equation}
    \Lambda>\Lambda_{\text{c,\ 3D}},
\end{equation}
where we have defined the critical value as a function of the gas parameter $\gamma$
\begin{equation}
     \Lambda_{\text{c,\ 3D}}\equiv\frac{1+\sqrt{1-z^2_0}\cos{\phi_0}}{\frac{z^2_0}{2}+\frac{16\sqrt{2}}{15\sqrt{\pi}} \sqrt{\gamma}\left[(1+z_0)^{\frac{5}{2}}+(1-z_0)^{\frac{5}{2}}-2\right]}.
\end{equation}
To understand the significance of the beyond-mean-field correction to the \gls{mqst} critical value one divides by the mean-field critical value $\Lambda_{\text{c,\ mf}}$
\begin{equation}
    \frac{\Lambda_{\text{c,\ 3D}}}{\Lambda_{\text{c,\ mf}}}=\Bigg[1+\frac{32\sqrt{2}}{15\sqrt{\pi}} \sqrt{\gamma}\frac{(1+z_0)^{\frac{5}{2}}+(1-z_0)^{\frac{5}{2}}-2}{z_0^2}\Bigg]^{-1}.
\end{equation}
Note that since the denominator of $\Lambda_{\text{c,\ 3D}}$ is larger than the $\Lambda_{\text{c,\ mf}}$ one, then the beyond-mean-field macroscopic quantum self-trapping critical value is smaller than the mean-field one  $\Lambda_{\text{c,\ 3D}}<\Lambda_{\text{c,\ mf}}$, as pictured in  Fig.\ref{fig:3DX}. Furthermore, the relative correction grows as the gas parameter decreases and it is marginally more significant for lower values of $|z_0|$.
\begin{figure}
\centering
\includegraphics[width=0.6\textwidth]{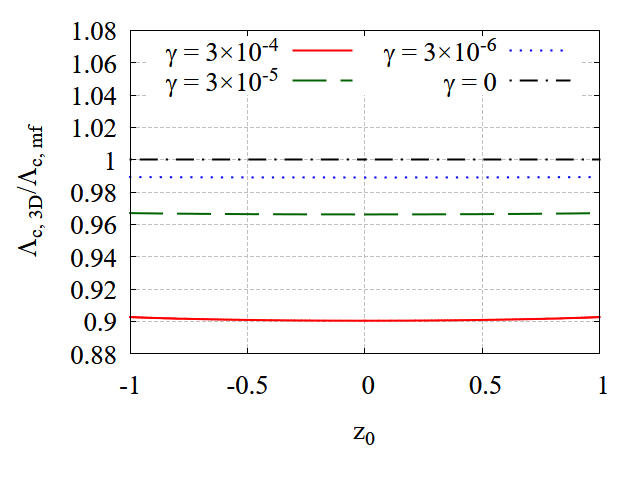}
\caption{3D beyond-mean-field relative correction to the \gls{mqst} critical value.\\ In the plot is pictured the ratio between the beyond-mean-field \gls{mqst} critical value $\Lambda_{\text{c,\ 3D}}$ and the mean-field one $\Lambda_{\text{c,\ mf}}$ as a function of the initial population imbalance $z_0\equiv z(t=0)=(n_1(0)-n_2(0))/(n_1(0)+n_2(0))$ for different values of the gas parameter $\gamma=a^3_sn$: $\gamma=3\times10^{-4}$ (red solid line),
$\gamma=3\times10^{-5}$ (green dashed line)
$\gamma=3\times10^{-6}$ (blue dotted line) and $\gamma=0$ (black dash-dotted line). The last line corresponds to the mean-field case.
}
\label{fig:3DX}
\end{figure}
\end{subsection}
\end{section}

\begin{section}{Beyond-mean-field: D=1 case}\label{sec:1d}
The procedure to find a modified Josephson frequency is analogous to the one used in the $D=3$ case.
{\cblue Recalling the procedure aforementioned, the beyond-mean-field energy density is derived from the renormalization of the integral over the momentum-space of the Bogoliubov spectrum, given by Eq.~(\ref{Bogoliubov}) with $D=1$.
Then, calculating the beyond-mean-field number density by taking a derivative of the beyond-mean-field grand potential with respect to the chemical potential and inverting it, an expression for the beyond-mean-field chemical potential as a function of the number density can be found. Finally, integrating this expression with respect to the number density allows you to obtain the beyond-mean-field energy density.
} Hence, from the $D=1$ beyond-mean-field energy density \cite{sala-zero,lieb}
\begin{equation}
        \mathcal{E}=\frac{1}{2}g_0n^2-\frac{2}{3\pi}\sqrt{\frac{m}{\hbar^2}}(g_0n)^{\frac{3}{2}},
\end{equation}
{\cblue where 
\begin{equation}
    g_0 = -\frac{2 \hbar^2 }{m a_s}.
\end{equation}
T}he Lagrangian density is given by
\begin{equation}
    \begin{split}
    \mathscr{L}=\mathscr{L}_0+\frac{2\sqrt{mg_0^3}}{3\pi\hbar}\left(|\Phi_1(t)|^3+|\Phi_2(t)|^3\right), 
    \end{split}
\end{equation}
where $\mathscr{L}_0$ is the mean-field Lagrangian density and the second term arises from the beyond-mean-field calculation accounting quantum fluctuation. 
The Lagrangian is thus
\begin{equation}
    \begin{split}
        \mathcal{L}&=\mathcal{L}_0+\frac{2\sqrt{mg_0^3}}{3\pi\hbar L^{\frac{1}{2}}}\left(|\varphi_1(t)|^3+|\varphi_2(t)|^3\right).
    \end{split}
\end{equation}
where the 1-dimensional coupling constant $g_0$ can be expressed as a function of the s-wave scattering length $a_s$ as follow
\begin{equation}
    g_0=-\frac{2\hbar^2}{ma_s}
\end{equation}
\begin{subsection}{Josephson frequency}
Again, a Madelung transformation (\ref{madelung}) is performed, obtaining 
\begin{equation}
        \mathcal{L}=\mathcal{L}_0+\frac{2\sqrt{mg_0^3}}{3\pi\hbar L^{\frac{1}{2}}}\left(N_1^{\frac{3}{2}}(t)+N_2^{\frac{3}{2}}(t)\right),
\end{equation}
which can be rewritten in terms of the total number of particles $N$ and the population imbalance $z$ as
\begin{equation}
    \mathcal{L}=\mathcal{L}_0+\frac{\sqrt{mg_0^3}N^{\frac{3}{2}}}{3\sqrt{2}\pi\hbar L^{\frac{1}{2}}}\left[(1+z)^{\frac{3}{2}}+(1-z)^{\frac{3}{2}}\right].
\end{equation}
Using the relation of expansion (\ref{expansion1}), valid in the low population imbalance limit, and removing the constant terms, one gets
\begin{equation}
    \mathcal{L}=\mathcal{L}_0+\frac{\sqrt{mg_0^3}N^{\frac{3}{2}}}{4\sqrt{2}\pi\hbar L^{\frac{1}{2}}}z^2.
\end{equation}
Finally, the Lagrangian in the $D=1$ case is given by
\begin{equation}
\begin{split}
            \mathcal{L}=&\frac{N\hbar}{2}z\Dot{\phi}-\left(\frac{UN^2+JN}{4}\right)z^2 -\frac{JN}{4}\phi^2+\frac{\sqrt{mg_0^3}N^{\frac{3}{2}}}{4\sqrt{2}\pi\hbar L^{\frac{1}{2}}}z^2,
\end{split}
\end{equation}
Therefore the Euler-Lagrange equations have the same form of Eq. (\ref{EL-freq}) but with the following corrected Josephson frequency
\begin{equation}
    \Omega\equiv\frac{1}{\hbar}\sqrt{J^2+JUN- \frac{J\sqrt{g_0^3nm}}{\sqrt{2}\pi\hbar}}.
\end{equation}
Writing now the Josephson frequency as a function of the s-wave scattering length $a_s$
\begin{equation}
    \Omega=\Omega_R\sqrt{1-\frac{2\hbar^2n}{ma_sJ}\left(1-\frac{1}{\pi\sqrt{-a_sn}}\right)},
\end{equation}
and defining the reference energy $\varepsilon_s$ and the gas parameter $\gamma$ in the 1-dimensional case as
\begin{equation}
    \varepsilon_s\equiv\frac{\hbar^2}{ma_s^2},\qquad\gamma\equiv a_sn,
\end{equation}
the Josephson frequency can also be written as 
\begin{equation}
    \Omega=\Omega_R\sqrt{1-2\gamma\frac{\varepsilon_s}{J}\left(1-\frac{1}{\pi\sqrt{-\gamma}}\right)}.
\end{equation}
Note that the gas parameter $\gamma$ must be negative due to the presence of the inverse of the square root of it.

Analogously to the 3-dimensional case to acknowledge the degree of the beyond-mean-field correction to the mean-field Josephson frequency is taking into account the ratio between the corrected frequency and the mean-field one as a function of the strength parameter $\Lambda=-2\gamma\varepsilon_s/J$
\begin{equation}\label{eq:freq-correction1d}
    \frac{\Omega}{\Omega_{mf}}=\sqrt{\frac{1+\Lambda\left(1-\frac{1}{\pi\sqrt{-\gamma}}\right)}{1+\Lambda}}.
\end{equation}
The beyond-mean-field Josephson frequency is lower than the mean-field one. Indeed, as pictured in Fig.~\ref{fig:1DO} the relative correction $\Omega/\Omega_{mf}\leq1$, where the equality is obtained for $\gamma\rightarrow-\infty$ or when the strength parameter is $\Lambda=0$. Analogously to the 3-dimensional case, in obtaining that beyond-mean-field correction, a perturbative analysis is used, so by setting a beyond-mean-field relative correction to the number density smaller than $0.1$, it follows that the upper bound on $\gamma$ must be set to $\gamma=-20$, as reported in Fig.~\ref{fig:1DO}.
Furthermore, the beyond-mean-field correction becomes more important as $\Lambda$ grows. In fact, for larger strength parameters, $\Lambda\gg1$, the asymptotic behavior of the relative correction is given by
\begin{equation}
    \frac{\Omega}{\Omega_{mf}}\Bigg|_{\Lambda\gg1}=\sqrt{1-\frac{1}{\pi\sqrt{-\gamma}}}.
\end{equation}
The relative correction $\Omega / \Omega_{mf}$ behavior also depends on the gas parameter $\gamma$, for higher values of the gas parameter the correction is more important and so the value of $\Omega/\Omega_{mf}$ decreases.
\begin{figure}
\centering
\includegraphics[width=0.6\textwidth]{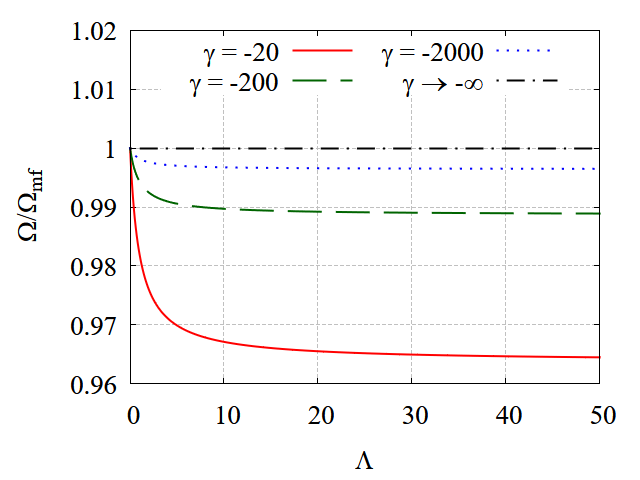}
\caption{1D beyond-mean-field relative correction to the Josephson frequency.\\ In the plot is pictured the ratio between the beyond-mean-field Josephson frequency $\Omega$ and the mean-field one $\Omega_ {mf}$ as a function of the strength parameter $\Lambda=g_0n/J$ for different values of the gas parameter $\gamma=a_sn$: $\gamma=-20$ (red solid line),
$\gamma=-200$ (green dashed line)
$\gamma=-2000$ (blue dotted line) and $\gamma\rightarrow-\infty$ (black dash-dotted line), which corresponds to the mean-field case.
}
\label{fig:1DO}
\end{figure}
\end{subsection}

\begin{subsection}{Macroscopic Quantum Self Trapping}
To compute the conserved energy one considers the beyond-mean-field Lagrangian in the $D=1$ case given by
\begin{equation}
\begin{split}
          \mathcal{L}=&\frac{N\hbar}{2}z\Dot{\phi}-\frac{UN^2}{4}z^2+\frac{JN}{2}\sqrt{1-z^2}\cos{\phi} +\frac{\sqrt{mg_0^3}N^{\frac{3}{2}}}{3\sqrt{2}\pi\hbar L^{\frac{1}{2}}}\left[(1+z)^{\frac{3}{2}}+(1-z)^{\frac{3}{2}}\right],
\end{split}
\end{equation}
one finds that the conserved energy is
\begin{equation}
\begin{split}
        E=&\frac{UN^2}{4}z^2-\frac{JN}{2}\sqrt{1-z^2}\cos{\phi}\\&-\frac{L\sqrt{m}}{3\sqrt{2}\pi\hbar}(UN)^{\frac{3}{2}}\left[(1+z)^{\frac{3}{2}}+(1-z)^{\frac{3}{2}}\right].
\end{split}
\end{equation}
Imposing the inequality condition to have \gls{mqst}, given by (\ref{INEQ}), one gets
\begin{equation}
\begin{split}
    &\frac{\Lambda}{2}z^2_0-\sqrt{1-z^2_0}\cos{\phi_0}-\frac{L\sqrt{2m}}{3\pi\hbar }\Lambda U^{\frac{1}{2}}N^{-\frac{1}{2}}\left[(1+z_0)^{\frac{3}{2}}+(1-z_0)^{\frac{3}{2}}\right] \\&>1-\frac{2L\sqrt{2m}}{3\pi\hbar }\Lambda U^\frac{1}{2}N^{-\frac{1}{2}},
\end{split}
\end{equation}
and finally
\begin{equation}
    \Lambda>\Lambda_{\text{c,\ 1D}},
\end{equation}
where we have defined the critical value $\Lambda_{\text{c,\ 1D}}$ using the gas parameter $\gamma$ 
\begin{equation}
\begin{split}
    \Lambda_{\text{c,\ 1D}} =& \Bigg(1+\sqrt{1-z^2_0}\cos{\phi_0}\Bigg)\Bigg[\frac{z^2_0}{2} -\frac{2}{3\pi}\frac{1}{\sqrt{-\gamma}} \left[(1+z_0)^{\frac{3}{2}}+(1-z_0)^{\frac{3}{2}}-2\right]\Bigg]^{-1}.
\end{split}
\end{equation}

However in this case the critical value is reached for larger values of $\Lambda$ since $\Lambda_{\text{c,\ 1D}}>\Lambda_{\text{c,\ mf}}$, namely the beyond-mean-field critical value $\Lambda_{\text{c,\ 1D}}$ is larger than the mean-field one. Indeed, dividing $\Lambda_{\text{c,\ 1D}}$ by the mean-field critical value $\Lambda_{\text{c,\ mf}}$ one gets
\begin{equation}
    \frac{\Lambda_{\text{c,\ 1D}}}{\Lambda_{\text{c,\ mf}}}=\Bigg(1-\frac{4}{3\pi} \frac{1}{\sqrt{-\gamma}}\frac{(1+z_0)^{\frac{3}{2}}+(1-z_0)^{\frac{3}{2}}-2}{z_0^2}\Bigg)^{-1},
\end{equation}
and looking at Fig.~\ref{fig:1DX} one finds $\Lambda_{\text{c,\ 1D}}/\Lambda_{\text{c,\ mf}}\geq1$, where the equality is verified for $\gamma\rightarrow-\infty$. In fact, the beyond-mean-field correction becomes less significant as the gas parameter decreases. Furthermore, at fixed $\gamma$  it is more important for higher values of $|z_0|$.
\begin{figure}
\centering
\includegraphics[width=0.6\textwidth]{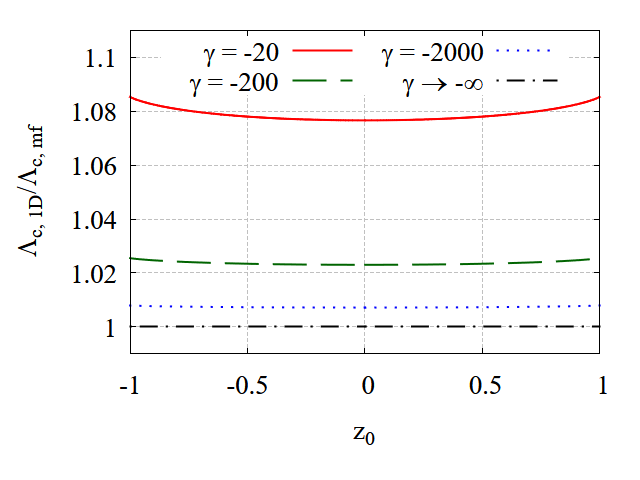}
\caption{1D beyond-mean-field relative correction to the \gls{mqst} critical value.\\
In the plot is pictured the ratio between the beyond-mean-field \gls{mqst} critical value $\Lambda_{\text{c,\ 1D}}$ and the mean-field one $\Lambda_{\text{c,\ mf}}$ as a function of the initial population imbalance $z_0\equiv z(t=0)=(n_1(0)-n_2(0))/(n_1(0)+n_2(0))$  for different values of the gas parameter $\gamma=a_sn$: $\gamma=-20$ (red solid line),
$\gamma=-200$ (green dashed line)
$\gamma=-2000$ (blue dotted line) and $\gamma\rightarrow-\infty$ (dark dash-dotted line). The last line corresponds to the mean-field case.
}
\label{fig:1DX}
\end{figure}
\end{subsection}
\end{section}

\begin{section}{Beyond-mean-field: D=2 case}\label{sec:2d}
The 2-dimensional case is more delicate that the 1-dimensional and the 3-dimensional cases for many reasons \cite{sala-zero,schick,popov}. Firstly, {\cblue renormalizing the zero-point energy of the excitations is not enough to deal with the UV divergency, since a $\Gamma(-2)$ appears. Under the minimal subtraction scheme we obtain a finite result \cite{lorenzi2023} for the zero-point energy of the excitations, which now depends on an energy cut-off $\Lambda$. However this energy cut-off cancels out with the same energy cut-off which appears in the expression for the mean-field 2-dimensional coupling constant, givin a $\Lambda-$indpendent expression for the beyond-mean-field grand potential. The beyond-mean-field grand potential is equal in form to the mean-field one, with the care of substituting the mean-field coupling constant with the beyond-mean-field one, namely:
\begin{equation}
    g_{r,k}=-\frac{4\pi\hbar^2}{m}\frac{1}{\ln{\left(\frac{\mu_k}{\varepsilon_0}\right)}}\qquad k=1,2 \ .
\end{equation}
Here, $\mu$ is the chemical potential and the constant $\varepsilon_0$ is given by $\varepsilon_0=4\hbar^2/(ma_s^2e^{2\gamma_E+\frac{1}{2}})$, where $a_s$
is the s-wave scattering length  and $\gamma_E=0.5772156649$ is the Euler-Mascheroni constant. We can then approximate this expression as a function of the} number density of the site as follows
\begin{equation}\label{eq:renorm}
    g_{r,k}=-\frac{4\pi\hbar^2}{m}\frac{1}{\ln{(Cn_k)}}\qquad k=1,2 \ ,
\end{equation}
where $C\equiv\pi e^{2\gamma_{\cblue E}+1}a^2_s$ . Secondly, the corrected Lagrangian is no more composed of a mean-field part and a correction, but rather it is equal in form to the mean-field Lagrangian although with the renormalized coupling $g_r$ replacing $g_0$.

Therefore, before computing the potential term, a discussion on the coupling is needed \cite{lorenzi2023}. Since the coupling $g_{k,r}$ is different for each site, it is useful to define a coupling $g_r$ for the entire system
\begin{equation}
    g_r=-\frac{4\pi\hbar^2}{m}\frac{1}{\ln{(Cn)}},
    \label{ga0}
\end{equation}
where $n$ is the number density of the system and it is given by the mean of the number densities of the sites
\begin{equation}
    n=\frac{n_1+n_2}{2}.
\end{equation}
Then the number densities of the sites $n_{1,2}$ can be expressed in terms of the population imbalance variable $z(t)=(n_1-n_2)/(n_1+n_2)$ and the  number density of the system $n$ as
\begin{equation}\label{imbalance}
\begin{split}
    n_1&=n(1+z) \\
    n_2&=n(1-z).
\end{split}
\end{equation}
The couplings $g_{r,k}$ in Eq.~(\ref{eq:renorm}) can rewritten in terms of $z(t)$, $n$ and $g_r$
and therefore, after some manipulations using Eq. (\ref{imbalance})
\begin{equation}
    g_{r,k}=g_r \Bigg[1+\frac{\ln{(1\pm z)}}{\ln{(Cn)}}\Bigg]^{-1}, \qquad k=1,2.
\end{equation}

\begin{subsection}{Josephson frequency}

In the case of $D=2$, the beyond-mean-field energy density is given by
\begin{equation}
    \mathcal{E}(n)=\frac{g_rn^2}{2}.
\end{equation}
Hence, the Lagrangian density is given by
\begin{equation}
\begin{split}
    \mathscr{L}=&\sum_{k}\left(i\hbar\Phi_k^*(t)\partial_t\Phi_k(t)-\frac{1}{2}g_{r,k}|\Phi_k(t)|^4\right) +\frac{J}{2}\left(\Phi_1^*(t)\Phi_2(t)+\Phi_2^*(t)\Phi_1(t)\right)
\end{split}
\end{equation}
and it is obtained substituting $g_0$ with $g_r$. Integrating in space the corresponding Lagrangian is
\begin{equation}
\begin{split}
    \mathcal{L}=&\sum_k\left(i\hbar\varphi^*_k(t)\partial_t\varphi_k(t)-\frac{U_k}{2}|\varphi_k(t)|^4\right)+\frac{J}{2}(\varphi_1^*(t)\varphi_2(t)+\varphi_2^*(t)\varphi_1(t)),
\end{split}
\end{equation}
where
\begin{equation}
    U_k\equiv\frac{g_{r,k}}{L^2},\qquad \varphi_k(t)\equiv L\Phi_k(t)\quad k=1,2.
\end{equation}
In the mean-field $D=2$ case we can easily express the potential terms as a function of the variables $N$ and $z$. This is because the coupling constant, denoted as $U$, is not influenced by the quantity $n_k$. Consequently, the potential term exhibits a quadratic dependence on $n_k$.
\begin{equation}
    -\sum_k\frac{U}{2}N^2_k=\frac{UN^2}{4}z^2.
\end{equation}
In the beyond-mean-field case it is not so simple since there is also a dependence on $n_k$ in the coupling $g_{r,k}$, hence the potential term transforms differently. For $k=1$, upon defining 
\begin{equation}
    U_r\equiv\frac{g_r}{L^2},
\end{equation}
and after performing a Taylor expansion due to the low population balance regime, $|z(t)|\ll1$, one finds
\begin{equation}
    \begin{split}
        \frac{1}{2}U_1|\varphi_1(t)|^4
        &\simeq\frac{U_rN^2}{8}\left(1+\frac{z^2-2z}{2\ln{(Cn)}}+\frac{z^2}{\ln^2{(Cn)}}\right)(1+2z+z^2),
    \end{split}
    \end{equation}
For $k=2$ the procedure is analogous.
Summing the two contributions one obtains
\begin{equation}
\begin{split}
\sum_k\frac{U_{r,k}}{2}|\varphi_k(t)|^4=&\frac{U_rN^2}{4}\Bigg(1+z^2 -\frac{3z^2}{2\ln{(Cn)}}+\frac{z^2}{\ln^2{(Cn)}}\Bigg).
\end{split}    
\label{urk}
\end{equation}
Writing this result in terms of the system coupling, inverting the relation (\ref{ga0})
\begin{equation}
    \frac{1}{\ln{(Cn)}}=-\frac{mg_r}{4\pi\hbar^2},
\end{equation}
one gets
\begin{equation}
    \begin{split}
        \sum_k\frac{U_{r,k}}{2}|\varphi_k(t)|^4&=\frac{U_rN^2}{4}\left[1+z^2\left(1+\frac{3}{8}\frac{mg_r}{\pi\hbar^2}+\frac{1}{16}\frac{m^2g_r^2}{\pi^2\hbar^4}\right)\right].
        \end{split}
\end{equation}
The term is similar to the mean-field one, given by  $(UN^2z^2)/4$ with caution to substitute $U$ with $U_r$ and add the beyond the mean-field corrections to the contact interaction term. Therefore to obtain the Josephson frequency in the 2-dimensional beyond-mean-field framework is sufficient to substitute inside (\ref{Omega}) the constant $U$ with
\begin{equation}
U\rightarrow U_r\left(1+\frac{3}{8}\frac{mg_r}{\pi\hbar^2}+\frac{1}{16}\frac{m^2g_r^2}{\pi^2\hbar^4}\right).
\label{Uraprox1}
\end{equation}
Doing so, the 2-dimensional beyond-mean-field Josephson frequency is given by
\begin{equation}
\begin{split}
    \Omega=&\frac{1}{\hbar}\sqrt{J^2+JU_rN\left(1+\frac{3}{8}\frac{mg_r}{\pi\hbar^2}+\frac{1}{16}\frac{m^2g_r^2}{\pi^2\hbar^4}\right)},
\end{split}
\end{equation}
or, alternatively, expressing it as a function of $g_r$ and $n$
\begin{equation}
    \Omega=\frac{1}{\hbar}\sqrt{J^2+Jg_rn\left(1-\frac{3}{2\ln{(Cn)}}+\frac{1}{\ln^2{(Cn)}}\right)}.
\end{equation}
Note that, in the limit of low density $n\ll1$, the terms $1/(\ln^\ell{(Cn)})$
become smaller and smaller the higher is $\ell$, therefore keeping only terms of the order $\ln({Cn})^{-1}$
then ($\ref{Uraprox1}$) can be approximated to $U_r$ and so the Josephson frequency in $D=2$  is formally equivalent to mean-field one (\ref{Omega})
\begin{equation}
    \Omega=\frac{1}{\hbar}\sqrt{J^2+U_rNJ},
\end{equation}
with the care of substituting $U$ with $U_r$.
Writing explicitly the Rabi frequency, the s-wave scattering length $a_s$ and the number density $n$ one obtains
\begin{equation}
    \Omega=\Omega_R\sqrt{1-\frac{4\pi\hbar^2n}{mJ\ln{(Cn)}}\left(1-\frac{3}{2\ln{(Cn)}}+\frac{1}{\ln^2{(Cn)}}\right)}.
\end{equation}
Introducing the reference energy $\varepsilon_s$ and the gas parameter in the 2-dimensional case
\begin{equation}
    \varepsilon_s\equiv\frac{\hbar^2}{ma^2_s},\qquad \gamma\equiv a_s^2n,
\end{equation}
and calling $C^*=\pi e^{2\gamma_{\cblue E}+1}$ the Josephson frequency can be written as
\begin{equation}
    \Omega=\Omega_R\sqrt{1-\frac{4\pi\gamma}{\ln{(C^*\gamma)}}\frac{\varepsilon_s}{J}\left(1-\frac{3}{2\ln{(C^*\gamma)}}+\frac{1}{\ln^2{(C^*\gamma)}}\right)}.
\end{equation}
The beyond-mean-field relative correction to the Josephson frequency $\Omega$ is given by
\begin{equation}
    \frac{\Omega}{\Omega_ {mf}}=\sqrt{\Bigg[1+\Lambda\left(1-\frac{3}{2\ln{(C^*\gamma)}}+\frac{1}{\ln^2{(C^*\gamma)}}\right)\Bigg]\Big(1+\Lambda\Big)^{-1}},
\end{equation}
where the strength parameter is given by $\Lambda=-(4\pi\gamma\varepsilon_s)/(\ln{(C^*\gamma)}J)$.\\
As pictured in Fig.~\ref{fig:2DO}, the relative $\Omega/\Omega_{mf}$ correction at fixed gas parameter $\gamma$ is more significant for higher values of the strength parameter $\Lambda$ and for larger values of the strength parameter $\Lambda$, the relative correction $\Omega/\Omega_{mf}$ is independent on $\Lambda$ and is given by
\begin{equation}
    \frac{\Omega}{\Omega_ {mf}}\Biggl|_{\Lambda\gg1}=\sqrt{1-\frac{3}{2\ln{(C^*\gamma)}}+\frac{1}{\ln^2{(C^*\gamma)}}}.
\end{equation}
Instead, focusing on the gas parameter $\gamma$ dependence one has an increment of the relative $\Omega/\Omega_{mf}$ correction for a higher value of $\gamma$. Conversely for $\gamma=0$ one retrieves the mean-field result.
\begin{figure}
\centering
\includegraphics[width=0.6\textwidth]{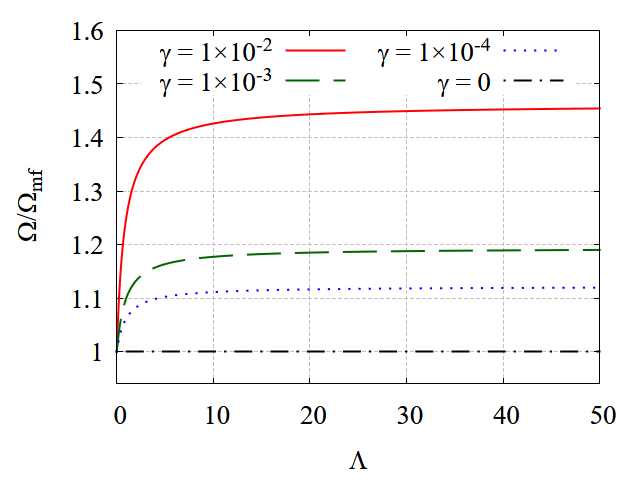}
\caption{2D beyond-mean-field relative correction to the Josephson frequency.
In the plot is pictured the ratio between the beyond-mean-field Josephson frequency $\Omega$ and the mean-field one $\Omega_ {mf}$ as a function of the strength parameter $\Lambda=g_0n/J$ for different values of the gas parameter $\gamma=a_s^2n$: $\gamma=1\times10^{-2}$ (red solid line),
$\gamma=1\times10^{-3}$ (green dashed line)
$\gamma=1\times10^{-4}$ (blue dotted line) and $\gamma=0$ (dark dash-dotted line). The last line corresponds to the mean-field case.
}
\label{fig:2DO}
\end{figure}
\end{subsection}

\begin{subsection}{Macroscopic Quantum Self Trapping} 
Unlike the Josephson frequency calculation, in the \gls{mqst} one, the low population imbalance limit is not taken. Therefore it is necessary to evaluate how the interaction terms transform in the case the population imbalance is generic.
Taking into account the contact interaction terms
\begin{equation}
    \begin{split}
        \sum_k\frac{1}{2}U_k|\varphi_k(t)|^4&=\frac{N^2}{8}[U_1(1+z)^2+U_2(1-z^2)].
    \end{split}
\end{equation}
Using the definitions of $U_r$ and $U_{k}$, they can be linked by the followed relation
\begin{equation}
    U_k=U_r \Bigg[1+\frac{\ln{(1\pm z)}}{\ln{(Cn)}}\Bigg]^{-1} \qquad k=1,2
\end{equation}
Then the interaction term reduces to
\begin{equation}
    \begin{split}
        \sum_k\frac{1}{2}U_k|\varphi_k(t)|^4 &=\frac{U_rN^2}{4}\Bigg[\Bigg(1+z^2+(1+z)^2\frac{\ln{(1- z)}}{2\ln{(Cn)}} +(1-z)^2\frac{\ln{(1+ z)}}{2\ln{(Cn)}}\Bigg) \\ &\times  \Bigg(\left(1+\frac{\ln{(1+ z)}}{\ln{(Cn)}}\right)\left(1+\frac{\ln{(1- z)}}{\ln{(Cn)}}\right)\Bigg)^{-1}\Bigg].
    \end{split}
\end{equation}

Hence, the beyond-mean-field Lagrangian in the $D=2$ case is
\begin{equation}
\begin{split}
     \mathcal{L}=&\frac{N\hbar}{2}z\Dot{\phi}+\frac{JN}{2}\sqrt{1-z^2}\cos{\phi} \\&-\frac{U_rN^2f(z)}{4}\Bigg(1+z^2 +\frac{(1+z)^2\ln{(1- z)}+(1-z)^2\ln{(1+z)}}{2\ln{(Cn)}}\Bigg),
\end{split}
\end{equation}
where the function $f(z)$ is introduced to lighten the expression 
\begin{equation}\label{eq:f}
    f(z)\equiv \Bigg[\left(1+\frac{\ln{(1+ z)}}{\ln{(Cn)}}\right)\left(1+\frac{\ln{(1- z)}}{\ln{(Cn)}}\right)\Bigg]^{-1}.
\end{equation}
From the Lagrangian one can compute the conserved energy
\begin{equation}
\begin{split}
     E&=\frac{U_rN^2f(z)}{4}\Big(1+z^2 +\frac{(1+z)^2\ln{(1- z)}+(1-z)^2\ln{(1+z)}}{2\ln{(Cn)}}\Big)\\&-\frac{JN}{2}\sqrt{1-z^2}\cos{\phi},
\end{split}
\end{equation}
and imposing the \gls{mqst} inequality condition, given by $E(z_0,\phi_0)>E(0,\pi)$, one obtains
\begin{equation}
\begin{split}
&\frac{U_rNf(z_0)}{2J}\Big(1+z_0^2+(1+z_0)^2\frac{\ln{(1-z_0)}}{2\ln{(Cn)}} +(1-z_0)^2\frac{\ln{(1+ z_0)}}{2\ln{(Cn)}}\Big) -\sqrt{1-z_0^2}\cos{\phi_0} \\&>1+\frac{U_rN}{2J}.
\end{split}
\end{equation}
\begin{figure}
\centering
\includegraphics[width=0.6\textwidth]{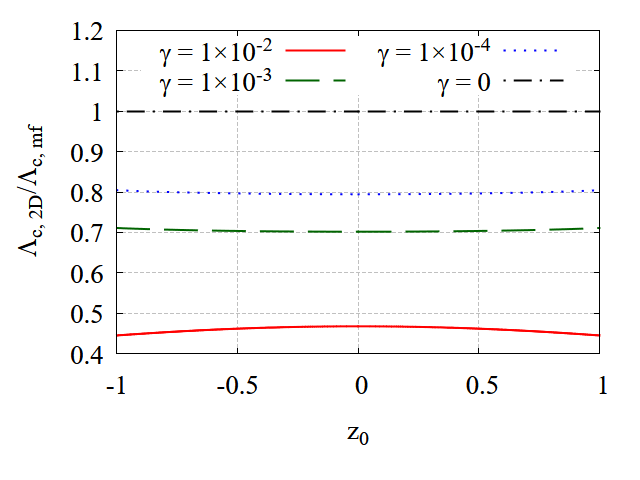}
\caption{2D beyond-mean-field relative correction to the \gls{mqst} critical value.\\ In the plot is pictured the ratio between the beyond-mean-field \gls{mqst} critical value $\Lambda_{\text{c,\ 2D}}$ and the mean-field one $\Lambda_{\text{c,\ mf}}$ as a function of the initial population imbalance $z_0\equiv z(t=0)=(n_1(0)-n_2(0))/(n_1(0)+n_2(0))$ for different values of the gas parameter $\gamma=a_s^2n$: $\gamma=1\times10^{-2}$ (red solid line),
$\gamma=1\times10^{-3}$ (green dashed line)
$\gamma=1\times10^{-4}$ (blue dotted line) and $\gamma=0$ (dark dash-dotted line). The last line corresponds to the mean-field case.
}
\label{fig:2DX}
\end{figure}
Defining the adimensional constant $\Lambda_r$ as
\begin{equation}
    \Lambda_r\equiv\frac{U_rN}{J},
\end{equation}
the inequality reduces to
\begin{equation}
    \Lambda_r>\Lambda_{\text{c,\ 2D}}
\end{equation}
where the critical value $\Lambda_{\text{c,\ 2D}}$ is given by the following definition, using
the gas parameter can be rewritten as
\begin{equation}
    \begin{split}
         \Lambda_{\text{c,\ 2D}}= \Bigg[1+\sqrt{1-z^2_0}\cos{\phi_0}\Bigg] \Bigg[\frac{z^2_0}{2}+\frac{(f(z_0)-1)(1+z_0^2)}{2} \\+\frac{f(z_0)\left[(1+z_0)^2\ln{(1-z_0)}+(1-z_0)^2\ln{(1+z_0)}\right]}{4\ln{(Cn)}} \Bigg]^{-1},
    \end{split}
\end{equation}
Dividing by the mean-field \gls{mqst} critical value $\Lambda_{\text{c,\ mf}}$ one obtains
\begin{equation}
    \begin{split}
         &\frac{\Lambda_{\text{c,\ 2D}}}{\Lambda_{\text{c,\ mf}}} = \Bigg[1+\frac{(f(z_0)-1)(1+z_0^2)}{z_0^2} \\&+\frac{f(z_0)\left[(1+z_0)^2\ln{(1-z_0)}+(1-z_0)^2\ln{(1+z_0)}\right]}{2\ln{(C^*\gamma)}z_0^2}\Bigg]^{-1}.
    \end{split}
\end{equation}
Looking at Fig.~\ref{fig:2DX} one notes that the ratio is equal to 1 when the gas parameter goes to zero, $\gamma=0$, retrieving the mean-field result. Since the beyond-mean-field correction is obtained by a perturbative analysis, by setting the maximum value of the relative correction to the number density to $0.1$, one obtains the upper bound on $\gamma$ of $\gamma<1\times 10^{-2}$, as reported in Fig.~\ref{fig:2DX}. Since in general the ratio is lower than 1, then one deducts that the beyond-mean-field correction decreases the \gls{mqst} critical value. Another consideration is that increasing the gas parameter the relative correction is more important since the ratio decreases.

\end{subsection}

\end{section}

{\cblue
\section{Solution of the equations of motion}\label{sec:slip}
\begin{figure}
    \centering
    \subfloat[$D=3$]{\includegraphics[width=0.49\textwidth]{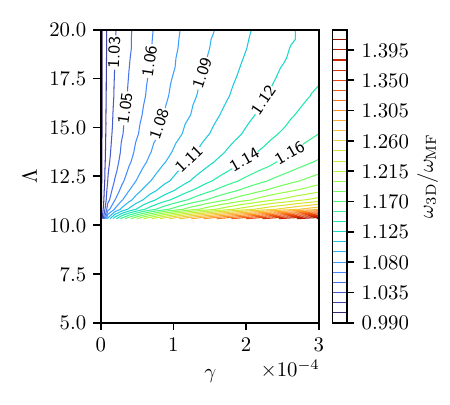} \label{fig:slip_3}}
    \subfloat[$D=2$]{\includegraphics[width=0.49\textwidth]{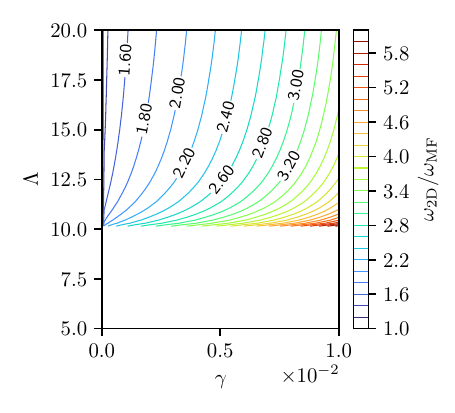} \label{fig:slip_2}}
    
    \subfloat[$D=1$]{\includegraphics[width=0.49\textwidth]{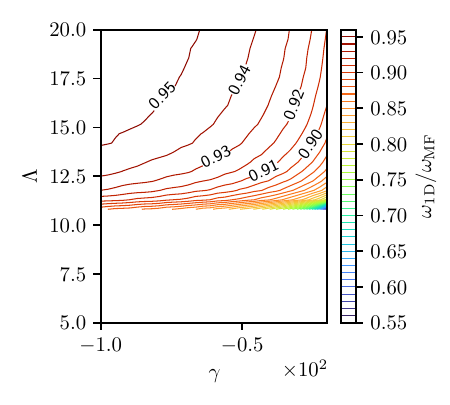} \label{fig:slip_1}}

    \caption{\cblue Comparison of the phase-slippage rate computed using the beyond-mean-field dynamics and the one computed using the mean-field dynamics.}
    \label{fig:slips}
\end{figure}

In this section, we solve numerically the equations of motion that are derived from the beyond-mean-field Lagrangians.
A particularly interesting dynamical feature to investigate is the phase-slippage rate \cite{avenel}, defined as
\begin{equation}
    \omega = \langle\dot{\phi}\rangle,
\end{equation}
where the average is taken over the whole evolution time, i.e. $\langle y\rangle \equiv \lim_{T\to \infty}\frac{1}{T} \int_{0}^{T} \dd{t} \ y(t)$.
This term depends on the difference between the chemical potentials. The chemical potential of the site $k$ is defined as $\mu_k = \pdv{E}{N_k}$, and the difference can be written as
\begin{equation}
    \Delta\mu = \mu_1 - \mu_2 = \pdv{E}{N_1} - \pdv{E}{N_2} = \frac{2}{N}\pdv{E}{z}
\end{equation}
This relation is exactly the formula for $\hbar \dot{\phi}$ coming from  Euler-Lagrange equations, or, equivalently, from Hamilton equation, so we can write  
\begin{equation}
    \dot{\phi} = \frac{\Delta \mu}{\hbar}
\end{equation}
This value can be nonzero in the regime of \gls{mqst} with running phase.

We write now the equations of motion, assuming for simplicity a scaled time $\tau = \hbar/J$. We also set the unit of mass to $m$, and the unit of action to $\hbar$.
In the mean-field case the equations become
\begin{subequations}
\begin{align}[left ={\empheqlbrace}]
    \dfrac{\dd{z}}{\dd{\tau}} &=-\sqrt{1-z^2}\sin{\phi} \label{motionMF1} \\
    \dfrac{\dd{\phi}}{\dd{\tau}} &=\dfrac{z}{\sqrt{1-z^2}}\cos{\phi}+\Lambda z.    \label{motionMF2}
\end{align}
\end{subequations}
It is clear that in the self-trapping regime Eq.~(\ref{motionMF2}) contains a steady state contribution to the derivative of the phase, due to the term $\Lambda z$, with $z$ that performs little oscillations around a nonzero mean value.
In the beyond-mean-field equations of motion, the equation for the population imbalance (\ref{motionMF1}) stays the same, and the one for the phase difference (\ref{motionMF2}), for $D=1, 3$, acquires some correction terms. The $D=3$ corrected equation reads
\begin{equation}
    \dfrac{\dd{\phi}}{\dd{\tau}} =\dfrac{z}{\sqrt{1-z^2}}\cos{\phi}+\Lambda z + \frac{8}{3} \sqrt{\frac{2}{\pi}} \ \Lambda \sqrt{\gamma}\left[ (1+z)^{3/2} - (1-z)^{3/2} \right],
\end{equation}
and in $D=1$ it reads 
\begin{equation}
    \dfrac{\dd{\phi}}{\dd{\tau}} =\dfrac{z}{\sqrt{1-z^2}}\cos{\phi}+\Lambda z - \frac{1}{\pi} \ \frac{\Lambda}{\sqrt{-\gamma}}\left[ (1+z)^{1/2} - (1-z)^{1/2} \right] .
\end{equation}
In the case of $D=2$, the equation changes form slightly and is written as
\begin{equation}
\begin{aligned}
    \dfrac{\dd{\phi}}{\dd{\tau}} = &\dfrac{z}{\sqrt{1-z^2}}\cos{\phi} + \frac{\Lambda_r}{4} \ln(C^* \gamma) \Bigg[(1+z)\frac{2\ln(C^*\gamma (1+z)) -1 }{\ln^2(C^* \gamma (1+z))} \\ & - (1-z)\frac{2\ln(C^*\gamma (1-z)) -1 }{\ln^2(C^* \gamma (1-z))} \Bigg].
\end{aligned}
\end{equation}
In Fig.~\ref{fig:slips} we consider the ratio of the phase-slippage rate in the beyond-mean-field and the mean-field dynamics, with initial conditions $z(0) = 0.6$ and $\phi(0) = 0$. In the $D=3$ case, represented in Fig.~\ref{fig:slip_3}, we observe that the frequency is enhanced for lower values of the strength and higher values of $\gamma$. At the same time, low values of $\gamma$ make the dynamics more similar to the mean-field case.
The case of $D=2$ is depicted in Fig.~\ref{fig:slip_2}, which reports a similar tendency as the $D=3$ case, but yielding significatively larger deviations.
Finally, the case of $D=1$ is represented in Fig.~\ref{fig:slip_1}. In this case, we notice that the frequency gets reduced by increasing $\gamma$, and has a weak decrease when decreasing $\Lambda$. We notice that the trend is opposite with respect to the other cases.
}
\section{Conclusions}
\begin{figure}
\centering
\includegraphics[width=0.6\textwidth]{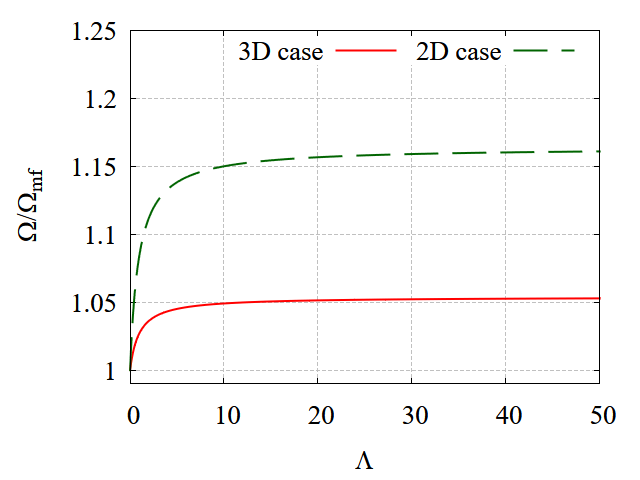}
\caption{\cblue{Comparison between the beyond-mean-field relative correction to the Josephson frequency in 3D and in 2D.
In the plot is pictured the ratio between the beyond-mean-field Josephson frequency $\Omega$ and the mean-field one $\Omega_ {mf}$ as a function of the strength parameter $\Lambda=g_0n/J$ for different dimensional cases: 3-dimensional case (red solid line),
2-dimensional case (green dashed line). The value of the gas parameter is $\gamma=3\times10^{-4}$ for each dimensional case.
}
\label{fig:32DO}
}
\end{figure}
\begin{figure}
\centering
\includegraphics[width=0.6\textwidth]{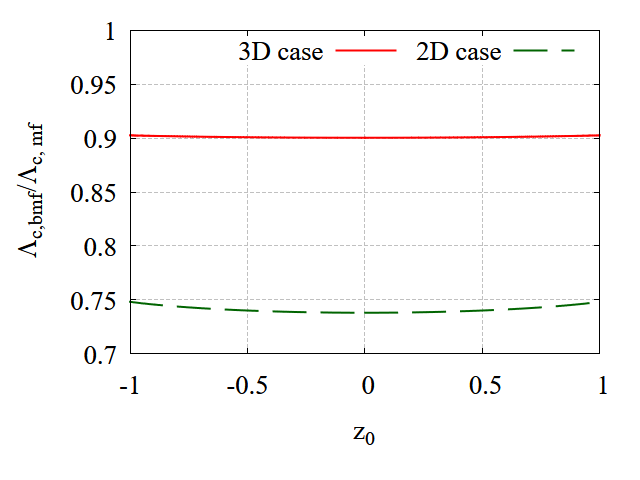}
\caption{\cblue{Comparison between the beyond-mean-field relative correction to the MQST critical value in 3D and in 2D.
In the plot is pictured the ratio between the beyond-mean-field MQST critical value $\Lambda_{c\; bmf}$ and the mean-field one $\Lambda_{c\; mf}$ as a function of the initial population imbalance $z_0\equiv z(t=0)=(n_1(0)-n_2(0))/(n_1(0)+n_2(0))$ for different dimensional cases at the same value of the gas parameter: 3-dimensional case (red solid line),
2-dimensional case (green dashed line). The value of the gas parameter is $\gamma=3\times10^{-4}$ for each dimensional case.
}
\label{fig:32DX}
}
\end{figure}

In this paper we have adopted the mean-field phase Lagrangian and applied Gaussian correction terms on both sites, taking into account a generic dimensionality of the system. Our calculations for $D=1, 2, 3$ have generalized the mean-field Josephson frequency in the low population-imbalance limit, and, on the other hand, computations for a severe population imbalance are provided for the critical \gls{mqst} strength. {\cblue The proposed method is complementary to the zero-dimensional approach described in Ref.~\cite{furutani2022quantum}. } The outcomes of our study indicate that, in situations where $D=2$ or $D=3$, the Josephson frequency is enhanced by the quantum corrections, whereas it is reduced when $D=1$. Moreover, we demonstrate that the critical strength for \gls{mqst} is lowered for $D=2$ or $D=3$ compared to mean-field calculations, while it is raised for $D=1$.
{\cblue In particular, as illustrated in Fig.~\ref{fig:32DO} and Fig.~\ref{fig:32DX} the corrections  in the $D=2$ case are stronger than the ones in the $D=3$ case when considering the same gas parameter. Focusing instead on the different behavior between the $D=3,2$ cases and the $D=1$ case could be explained as follows: in all the three-dimensional cases the ratio between the beyond-mean-field Josephson frequency and the mean-field one increases when the absolute value of the gas parameter increases. The latter can be interpreted as the number of particles inside a $D$-dimensional box of size given by the $D$-dimensional s-wave scattering length $a_s$. As $\gamma$ increases, the system is denser, and thus the Josephson frequency is enhanced in all the cases. However, while in the $D=3,2$ cases the mean-field approach is an adequate approximation for highly dilute systems, or in other words when the gas parameter is close to zero; viceversa, in the $D=1$ case the mean-field approximation is correct once one considers systems with the limit of the gas parameter that is going to $-\infty$. This difference is related to the expression of the interaction parameter $g_0$, which is qualitatively different in the various dimensions \cite{lorenzi2023}. Indeed, while in $D=3$ we have $g_0 \propto  a_s$, in $D=2$ we have $g_0 \propto \ln^{-1}(\kappa a_s)$, with a constant $\kappa$, and in $D=1$, the relation is $g_0 \propto a_s^{-1}$.  To summarise, the values of the corrections for each dimension are shown in Tab.~\ref{tab:tab3} and Tab.~\ref{tab:tab4}.
\begin{table}
	         \centering
	         \hspace{10cm}\begin{tabular}{|c|c|}
          \hline
	             D &  $\Omega/\Omega_{\mathrm{mf}}$  \\
              \hline
	     3  &  $\sqrt{\frac{1+\Lambda\left(1+8\sqrt{\frac{2\gamma}{\pi}}\right)}{1+\Lambda}}$\\       2 & $\sqrt{\frac{1+\Lambda\left(1-\frac{3}{2\ln{(C^*\gamma)}}+\frac{1}{\ln^2{(C^*\gamma)}}\right)}{1+\Lambda}}$ \\[1pt]
             1 & $\sqrt{\frac{1+\Lambda\left(1-\frac{1}{\pi\sqrt{-\gamma}}\right)}{1+\Lambda}}$ \\[1pt]        
             \hline
	         \end{tabular}
	         \hspace{-2cm}\caption{{\cblue Beyond-mean-field relative corrections to the Josephson frequency in dimension $D=1,2,3$. The correction is determined by the $D$-dimensional gas parameter $\gamma\equiv a_s^Dn$, where $a_s$ represents the s-wave scattering length $a_s$, and $n$ is the number density, and the strength parameter $\Lambda\equiv UN/J$. In the case of $D=2$, the result depends also on the parameter $C^*=\pi e^{2\gamma_E+1}a^2_s$, where $\gamma_E=0.5772156649$ is the Euler-Mascheroni constant.}}
	         \label{tab:tab3}
	     \end{table}
 \begin{table}
	         \centering
	         \hspace{10cm}\begin{tabular}{|c|c|}
          \hline
	             D &  $\Lambda_{\mathrm{c,\;bmf}}/\Lambda_{\mathrm{c,\;mf}}$  \\
            \hline
                        
            3  & $\Bigg(1+\frac{32\sqrt{2}}{15\sqrt{\pi}} \sqrt{\gamma}\frac{(1+z_0)^{\frac{5}{2}}+(1-z_0)^{\frac{5}{2}}-2}{z_0^2}\Bigg)^{-1}$
            \\ 
                        
            2 & $\!\begin{aligned}
            &\Big(1+\textstyle\frac{(f(z_0)-1)(1+z_0^2)}{z_0^2} \\ &+ \textstyle\frac{f(z_0)\left[(1+z_0)^2\ln{(1-z_0)}+(1-z_0)^2\ln{(1+z_0)}\right]}{2\ln{(C^*\gamma)}z_0^2}\Big)^{-1}
            \end{aligned}$
            \\
                    
            1  & $\Bigg(1-\frac{4}{3\pi} \frac{1}{\sqrt{-\gamma}}\frac{(1+z_0)^{\frac{3}{2}}+(1-z_0)^{\frac{3}{2}}-2}{z_0^2}\Bigg)^{-1}$\\   
            
            \hline
	         \end{tabular}
	         \hspace{-2cm}\caption{{\cblue Beyond-mean-field relative corrections to the MQST critical value in dimension $D=1,2,3$. The correction is determined by the $D$-dimensional gas parameter $\gamma\equiv a_s^Dn$, where $a_s$ represents the s-wave scattering length $a_s$, and $n$ is the number density, and on the initial population imbalance $z_0\equiv z(t=0)=(n_1(0)-n_2(0))/(n_1(0)+n_2(0))$. In the case of $D=2$, the result depends also on the parameter $C^*=\pi e^{2\gamma_E+1}a^2_s$, where $\gamma_E=0.5772156649$ is the Euler-Mascheroni constant and on the function $f(z)$ reported in Eq. (\ref{eq:f}).}}
	         \label{tab:tab4}
	     \end{table}
Our results suggest a method to verify experimentally the effects of beyond-mean-field corrections. For example let us consider an elongated bosonic Josephson junction in one dimension, made with a trapped gas of $^{87}$Rb, similar to a recent experiment \cite{pigneur}. Consider a hopping term of $J/h = 2 \  $Hz, with $h$ the Planck constant, and with the same trap frequencies of the reported setup. Let us suppose the trap frequency such that the axial confinement length is $l_z = \sqrt{\hbar /(m\omega_z)} \approx 1000 \ \mu$m, and the transverse one is $l_\perp = \sqrt{\hbar /(m\omega_\perp)} \approx 0.2 \ \mu$m.
Initiating the atom populations to a total atom number $N=41000$ with a low imbalance, and setting the three-dimensional scattering length to $a_{s, \mathrm{3D}} = 1600 a_0$ with $a_0$ the Bohr radius \cite{notarella, effective}, we predict the measure of the Josephson frequency with a deviation from the mean-field value by a reduction of about $3.5\%$.}
We believe that our theoretical predictions could be tested in near-future experiments with superfluid Josephson junctions made of ultracold atomic quantum gases. 

\section*{Acknowledgements}
{\cblue FL thanks Andrea Richaud for useful comments and suggestions.}
This work is partially supported by the BIRD grant “Ultracold atoms in curved
geometries” of the University of Padova, by the “IniziativaSpecifica Quantum” of INFN, by the European Union-NextGenerationEU within the National Center for HPC, Big Data and Quantum Computing (Project No. CN00000013, CN1 Spoke 10: “Quantum Computing”),  by the EU Project PASQuanS 2 ``Programmable Atomic Large-Scale Quantum Simulation"{\cblue, and the National Grant of the Italian Ministry of University and Research for the PRIN 2022 project "Quantum Atomic Mixtures: Droplets, Topological Structures, and Vortices".}


\end{document}